\documentstyle[12pt,subeqn]{article}

\newcommand\vek[1]{\mbox{\rmfamily\bfseries\itshape#1}}

\def\rd{{\rm d}}

\begin{document}

\title{Surface Tension of an Ideal Dielectric-Electrolyte 
Boundary: Exactly Solvable Model}

\author{
L. {\v S}amaj$^*$
}

\maketitle

\noindent Laboratoire de Physique Th{\'e}orique, Universit{\'e}
de Paris-Sud, B{\^a}timent 210, 91405 Orsay Cedex, France (Unit{\'e}
Mixte de Recherche no. 8627 - CNRS);
e-mail: samaj@th.u-psud.fr

\vskip 0.5truecm

\begin{abstract}
The model under consideration is a semi-infinite two-dimensional
two-component plasma (Coulomb gas), stable against bulk collapse for 
the dimensionless coupling constant $\beta<2$, in contact with
a dielectric wall of dielectric constant $=0$.
The model is mapped onto an integrable sine-Gordon theory
with a ``free'' Neumann boundary condition.
Using recent results on a reflection relationship between 
the boundary Liouville and sine-Gordon theories, an explicit 
expression is derived for the surface tension at a rectilinear
dielectric -- Coulomb gas interface.
This expression reproduces the Debye-H\"uckel $\beta \to 0$ limit
and the exact result at the bulk collapse border, the free-fermion 
point $\beta =2$, where the surface tension keeps a finite value.
The surface collapse, identified with the divergence of the
surface tension, occurs at $\beta =3$.
\end{abstract}


\medskip
\noindent LPT Orsay 00-134

\vfill

\noindent $^*$ On leave from the Institute of Physics, Slovak Academy
of Sciences, Bratislava, Slovakia

\newpage

\section{Introduction}
In a previous paper \cite{Samaj1},
the bulk thermodynamic properties (free energy, specific heat,
etc\ldots) of the two-dimensional (2D) two-component plasma
(with logarithmic interactions), or Coulomb gas, 
have been obtained exactly in the whole temperature
range for which the point-particle model is stable against collapse, 
i.e., for the dimensionless coupling constant 
(inverse temperature) $\beta<2$. 
The mapping onto a bulk sine-Gordon field theory was made and
recent results about that field theory \cite{Zamolodchikov},
\cite{Lukyanov}, were used.
In a next paper \cite{Samaj2},
a surface property of the same model in contact with an ideal
conductor of dielectric constant $\to\infty$, impenetrable
to particles, was considered.
In particular, the surface tension at a rectilinear conductor -- 
Coulomb gas interface was obtained exactly as a function 
of the bulk density, the applied potential and the temperature.
A mapping onto an integrable boundary sine-Gordon theory
\cite{Ghoshal} with a Dirichlet boundary condition was 
established, and known results about that field theory
\cite{Skorik}, \cite{LeClair}, were used.

The model considered in the present paper is the 2D Coulomb gas in
contact with an ideal dielectric wall of dielectric constant $=0$.
The system is mapped onto an integrable sine-Gordon theory,
now with a ``free'' Neumann boundary condition.
Using very recent results on a ``reflection'' relationship
between the boundary Liouville and sine-Gordon theories
\cite{Fateev1} - \cite{Fateev3}, an explicit expression is
derived for the surface tension at a rectilinear 
dielectric -- Coulomb gas interface as a function of the bulk
density, or the fugacity, and the temperature.
The surface tension is checked on its high-temperature
expansion derived from a renormalized Mayer expansion and
on the exact result at the bulk collapse border, the
free-fermion point $\beta = 2$, where it keeps a finite
value \cite{Jancovici1}.
The divergency of the surface tension occurs at $\beta = 3$.  

The model under consideration mimics the interface between
an electrolyte (the 2D two-component plasma, made of two
species of point particles, of opposite charges $\pm 1$)
and an ideal dielectric wall.
Classical equilibrium statistical mechanics is used.
In the grand-canonical formalism, the control parameters
are the inverse temperature $\beta$ and the two fugacities
$z_+$ and $z_-$ of the positive and negative particles, 
respectively. 
Due to the charge neutrality, 
the bulk properties of the plasma depend only on the
combination $z = (z_+ z_-)^{1/2}$ \cite{Lieb}.
It is a specific feature of the present model that also its 
surface properties depend only on $z$.

In the bulk of the plasma, the interaction Boltzmann factor 
of a positive-negative pair of charges, $r^{-\beta}$, is integrable 
at small $r$ if and only if $\beta<2$ (at large $r$, the interaction
is screened), with $\beta = 2$ being the bulk collapse
border of the plasma.
In the region $\beta \ge 2$, one has to attach to particles 
a small hard core $\sigma$ in order to prevent the collapse.
As has been argued long time ago \cite{Hauge}, in the limit 
$\sigma \to 0$, while the free energy and the internal energy 
per particle diverge, the specific heat, truncated correlation
functions, etc. remain finite.
In this sense one should also consider our exact result
for the surface tension when $\beta \ge 2$.
The stability range for surface properties of the model
is $\beta < 3$.
The surface collapse at $\beta = 3$ is identified with 
the divergency of the surface tension.

Series expansions and exact results are rather rare for
the present 2D configuration electrolyte -- dielectric wall.
More success was attained in the case when the electrolyte
is modeled by a one-component plasma, i.e., a system of
moving point particles of the same charge embedded in a rigid
neutralizing background.
The $\beta \to 0$ limit was treated in ref. \cite{Alastuey}
and the $\beta=2$ free fermion case was solved by Smith \cite{Smith}
(thermodynamics) and Jancovici \cite{Jancovici2} (density
profile, pair correlations, sum rules).
When the electrolyte is modeled by a two-component plasma,
at $\beta = 2$, the exact solution was available only for a hard wall
and an ideal conductor wall \cite{Cornu}, \cite{Forrester},
which admit a Thirring representation with an inhomogeneous mass 
going to 0 and to $\infty$ within the wall, respectively.
Only recently \cite{Jancovici1}, an exact solution has been obtained 
also for the dielectric wall of interest by the method of pfaffians,
and that solution offers via an explicit density profile a check of 
the surface tension formula at $\beta = 2$.

One should emphasize that the present result was
obtained due to recent progress in integrable field theories,
in particular, a reflection relationship between the boundary
Liouville and sine-Gordon theories \cite{Fateev1} - \cite{Fateev3}.
Mainly the work \cite{Fateev3}, providing an exact formula for
the one-point function of an exponential boundary operator
in the boundary sine-Gordon model with a general integrable
boundary action, was crucial.
The surface tension is obtained as a result of an integration from
the conductor wall (Dirichlet boundary condition) with a
known formula for the surface tension \cite{Samaj2}, to the 
dielectric wall (Neumann boundary condition), via a continuous
class of integrable boundary field theories.

The paper is organized as follows.
In Section 2, the mapping of the Coulomb gas in contact with an
ideal dielectric onto a sine-Gordon field theory with a Neumann
boundary condition is made.
The reflection relationship between the boundary Liouville 
and sine-Gordon theories is briefly explained and basic
formulae are written down in Section 3.
The exact formula for the surface tension is derived 
in Section 4.
Its high temperature expansion and the $\beta=2$ solution
are checked in Section 5.
Concluding remarks about a possible mechanism for the
surface collapse are given in Section 6.

\section{Mapping}
We consider an infinite 2D space of points ${\vek r} \in R^2$
defined by Cartesian coordinates $(x,y)$.
The model interface is localized at $x=0$, along the $y$ axis.
The half-space $x<0$, impenetrable to particles, is assumed 
to be occupied by an ideal dielectric of dielectric constant $=0$.
The electrolyte in the complementary half-space $x>0$ is
modeled by the classical 2D TCP of point particles $\{ j \}$
of charge $\{ q_j = \pm 1 \}$, immersed in a homogeneous 
medium of dielectric constant $=1$.
Classical equilibrium statistical mechanics is used.
In the grand-canonical formalism, the control parameters
are the inverse temperature $\beta$ and the fugacities
$z_+$ and $z_-$ of the positive and negative particles,
respectively.
For the dielectric wall of interest, only strictly neutral
charge configurations survive and, therefore, the thermodynamics 
depends only on $z = (z_+ z_-)^{1/2}$ as will follow directly
from the formalism presented here.
Due to the translational invariance in the $y$ direction,
the particle densities $n_+({\vek r}) = n_-({\vek r})$
depend only on $x$.
Let us denote their asymptotical $x\to\infty$ values by
$n_+ = n_- = n/2$ where $n$ is the total particle number
density.

The interaction energy $E$ of particles $\{ q_j,
{\vek r}_j = (x_j>0,y_j)\}$ consists of two parts
(see, e.g., \cite{Jackson}),
\begin{equation} \label{1}
E = \sum_{i<j} q_i q_j v(\vert {\vek r}_i - {\vek r}_j \vert)
+ {1\over 2} \sum_{i,j} q_i q_j 
v(\vert {\vek r}_i - {\vek r}_j^* \vert)
\end{equation}
where ${\vek r}^* = (-x,y)$ and $v({\vek r}) = 
- \ln(\vert r\vert /r_0)$ is the 2D Coulomb potential 
($r_0$ will be set for simplicity to unity). 
The first term corresponds to direct particle-particle
interactions, while the second one to interactions of
particles with the images due to the presence of 
the dielectric wall.
Introducing the microscopic charge + image charge density
\begin{equation} \label{2}
{\tilde \rho}({\vek r}) = \sum_j q_j \left[ \delta(x-x_j)
+ \delta(x+x_j) \right] \delta(y-y_j)
\end{equation} 
the energy (\ref{1}) can be written as
\begin{equation} \label{3}
E = {1\over 4} \int \rd^2 r \int \rd^2 r' ~ 
{\tilde \rho}({\vek r}) v({\vek r},{\vek r}')
{\tilde \rho}({\vek r}') - {1\over 2} N v(0) 
\end{equation}
where the integrations over ${\vek r}$ and ${\vek r}'$ are 
taken over the whole 2D space and $-(N/2)v(0)$ is the self-energy.

The grand-canonical partition function is defined by
\begin{subequations} \label{4}
\begin{equation} \label{4a}
\Xi = \sum_{N_+ =0}^{\infty} \sum_{N_- =0}^{\infty}
{z_+^{N_+}\over N_+!} {z_-^{N_-}\over N_-!} Q(N_+,N_-) 
\end{equation}
where
\begin{equation} \label{4b}
Q(N_+,N_-) = \int \prod_{j=1}^{N} \rd^2 r_j 
\exp \left[ -\beta E(\{ q_j,{\vek r}_j\})\right]
\end{equation}
\end{subequations}
is the canonical partition function of $N_+$ positive and
$N_-$ negative charges and $N = N_+ + N_-$.
Since, in infinite space, $-\Delta/(2\pi)$ is the inverse 
operator of $v({\vek r})$, the Boltzmann factor of the interaction
energy is expressible as
\begin{equation} \label{5}
\exp \left[ - {\beta\over 4} \int \rd^2 r \int \rd^2 r'
{\tilde \rho}({\vek r}) v({\vek r},{\vek r}') {\tilde \rho}({\vek r}')
\right] = {\int {\cal D} \phi \exp \left[ \int \rd^2 r \left(
{1\over 2} \phi \Delta \phi + {\rm i} \sqrt{\pi\beta}\phi
{\tilde \rho} \right) \right] \over \int {\cal D} \phi
\exp \left( \int \rd^2 r {1\over 2} \phi \Delta \phi \right)}
\end{equation} 
where $\phi({\vek r})$ is a real scalar field and $\int {\cal D}
\phi$ denotes the functional integration over this field.
Inserting ${\tilde \rho}$ from (\ref{2}), one recognizes in
the action of the field theory (\ref{5}) a nonlocal term
${\rm i} \sqrt{\pi\beta} \sum_j q_j [\phi(x_j,y_j) + \phi(-x_j,y_j)]$.
To make the field theory local, we shall reformulate it as
a boundary problem by introducing two new fields \cite{Callan}
\begin{subequations} \label{6}
\begin{eqnarray} 
\phi_e(x,y) & = & {1\over \sqrt{2}} \left[ \phi(x,y) + \phi(-x,y)
\right] \label{6a} \\
\phi_o(x,y) & = & {1\over \sqrt{2}} \left[ \phi(x,y) - \phi(-x,y)
\right] \label{6b}
\end{eqnarray}
\end{subequations}
defined only in the positive $x\ge 0$ half-space.
The even field has a Neumann boundary condition 
$\partial_x \phi_e \vert_{x=0} = 0$ and the odd field has a 
Dirichlet boundary condition $\phi_o \vert_{x=0} = 0$.
Since it holds
\begin{equation} \label{7}
\int \rd^2 r {1\over 2} \phi \Delta \phi = {1\over 2} 
\int_{x>0} \rd^2 r \left( \phi_e \Delta \phi_e +
\phi_o \Delta \phi_o \right)
\end{equation}
the odd field, contributing only by its free-field
part $\phi_o\Delta \phi_o/2$, disappears from (\ref{5})
by the numerator-denominator cancelation.
By integration per partes, the term $\phi_e \Delta \phi_e$
can be rewritten as $-(\nabla \phi_e)^2$, with a vanishing
contribution from the boundary.
The rhs of (\ref{5}) is thus expressible as
\begin{equation} \label{8}
{\int {\cal D} \phi_e \exp \left\{ \int_{x>0} \rd^2 r
\left[ - {1\over 2} (\nabla \phi_e)^2 \right] + {\rm i} \sqrt{2\pi \beta}
\sum_j q_j \phi_e({\vek r}_j) \right\} \over
\int {\cal D} \phi_e \exp \left[ - \int_{x>0} \rd^2 r {1\over 2}
(\nabla\phi_e)^2 \right]}
\end{equation} 
After inserting (\ref{8}) via (\ref{5}) into (\ref{4}),
one proceeds along the standard line \cite{Samaj2} and finally 
expresses $\Xi$ in terms of the 2D Euclidean sine-Gordon theory 
formulated in the half-space $x>0$:
\begin{equation} \label{9}
\Xi = {\int {\cal D} \phi \exp \left\{ \int_{x>0} \rd^2 r
\left[ - {1\over 2} (\nabla \phi)^2 + 2 z \cos 
(\sqrt{2\pi\beta}\phi) \right] \right\} \over
\int {\cal D} \phi \exp \left[ - \int_{x>0} \rd^2 r {1\over 2}
(\nabla\phi)^2 \right]}
\end{equation}
with Neumann condition $\partial_x \phi(x,y) \vert_{x=0} = 0$ 
for the field at the boundary.
Here, $z$ is the fugacity $(z_+ z_-)^{1/2}$ renormalized by 
the self-energy term $\exp[\beta v(0)/2]$, and the uniformly 
shifted $\phi_e$ is renamed as $\phi$,
\begin{equation} \label{10}
\phi = \phi_e + {\ln(z_+/z_-) \over 2 {\rm i} \sqrt{2\pi \beta}}
\end{equation}
The dependence of the statistics exclusively on
$(z_+ z_-)^{1/2}$ is now evident.
The rescaling of the field $\phi \to \phi/\sqrt{2\pi}$ transforms
(\ref{9}) into the form
\begin{subequations} \label{11}
\begin{equation} \label{11a}
\Xi(z) = {\int {\cal D} \phi ~ \exp [ - {\cal A}_{sG}(z)] \over
\int {\cal D} \phi ~ \exp [ - {\cal A}_{sG}(z=0)]}
\end{equation}
with the action
\begin{eqnarray} 
{\cal A}_{sG}(z) & = & \int_{x>0} {\rm d}^2 r \left[
{1\over 4 \pi} (\nabla \phi)^2 - 2 z \cos (2 b \phi) \right]
\label{11b} \\
b & = & {1\over 2} \sqrt{\beta} \label{11c}
\end{eqnarray}
\end{subequations}
and boundary condition $\partial_x \phi \vert_{x=0} = 0$,
which is more convenient for our purpose.
Hereinafter, the dependence of quantities on the temperature
parameter $b$ will be omitted in the notation.

The boundary sine-Gordon theory (\ref{11}) is a member of 
integrable sine-Gordon theories defined in the half-space
$x>0$ \cite{Ghoshal}:
\begin{subequations} \label{12}
\begin{equation} \label{12a}
\Xi(z,z_B) = {\int {\cal D} \phi ~ \exp [ - {\cal A}_{sG}(z,z_B)] \over
\int {\cal D} \phi ~ \exp [ - {\cal A}_{sG}(z=0,z_B)] }
\end{equation}
with the action
\begin{equation} \label{12b}
{\cal A}_{sG}(z,z_B)  =  \int_{x>0} {\rm d}^2 r \left[
{1\over 4 \pi} (\nabla \phi)^2 - 2 z \cos (2 b \phi) \right]
- 2 z_B \int_{-\infty}^{\infty} \rd y ~ \cos (b \phi_B)
\end{equation}
\end{subequations}
Here, $\phi_B(y) = \phi(x=0,y)$ is the boundary field and
$z_B$ the boundary fugacity.
The underlying sine-Gordon theory (\ref{11}) with Neumann
boundary condition $\partial_x \phi \vert_{x=0} = 0$ is known to
correspond to the ``free'' case $z_B=0$
(see, e.g., \cite{Ghoshal}, \cite{Fendley}).
The model of the metal-electrolyte boundary studied in 
ref. \cite{Samaj2} corresponds to the limit $z_B \to \infty$,
fixing the value of the boundary field $\phi \vert_{x=0} = 0$
(Dirichlet boundary conditions for zero potential difference 
between the metal and the electrolyte).

The grand potential $\Omega = - \beta^{-1} \ln \Xi(z,z_B)$
is the sum of a volume part and a surface part:
\begin{equation} \label{13}
\Omega = - V p(z) + S \gamma(z,z_B)
\end{equation} 
where $p$ is the bulk pressure, dependent only on $z$, and
$\gamma$ the surface tension, dependent on both $z$ and $z_B$.
For a strip $L \times R$, $R \to \infty$ in the $y$ direction and
$L$ large in the $x$ direction, the ``specific'' $\Omega/R$
is given by
\begin{equation} \label{14}
\lim_{R\to\infty} {\Omega \over R} = - L p(z) + \gamma(z,z_B)
\end{equation}
Thus, with regard to (\ref{12}) and taking into account the
crucial $z_B$-independence of $p$, one finds
\begin{equation} \label{15}
{\partial \over \partial z_B} \beta \gamma (z,z_B) = - 2
\left[ \langle {\rm e}^{{\rm i}b\phi_B} \rangle_{z,z_B} -
\langle {\rm e}^{{\rm i}b\phi_B} \rangle_{z=0,z_B} \right]
\end{equation}
where the obvious symmetry relation
\begin{equation} \label{16}
\langle {\rm e}^{{\rm i}b\phi_B} \rangle_{z,z_B} =
\langle {\rm e}^{-{\rm i}b\phi_B} \rangle_{z,z_B}
\end{equation}
implied by the invariance of the action (\ref{12b}) with respect
to the transformation $\phi \to -\phi$, was assumed.
$\langle \ldots \rangle_{z,z_B}$ denotes the averaging with
the sine-Gordon action (\ref{12b}).

\section{Reflection property}
The Liouville field theory, formulated in the half-space $x>0$
with a conformally invariant condition at $x=0$, is defined
by the action
\begin{equation} \label{17}
{\cal A}_{Liouv}(z,z_B) = \int_{x>0} \rd^2 r \left[ {1 \over 4\pi}
(\nabla \phi)^2 + z {\rm e}^{2b\phi} \right] + 
z_B \int_{-\infty}^{\infty} \rd y ~ {\rm e}^{b\phi_B}
\end{equation}
and the boundary condition on the field $\phi$ at infinity,
$\phi(x,y) = - Q \ln (x^2) + O(1)$ as $x\to\infty$.
The quantity
\begin{equation} \label{18}
Q = b + 1/b
\end{equation}
is called the ``background charge''.
The two-point function of the exponential of the boundary field
behaves like
\begin{equation} \label{19}
\langle {\rm e}^{a \phi_B(y)} {\rm e}^{a \phi_B(y')}
\rangle_{Liouv} = {d(a\vert z,z_B) \over \vert y - y' 
\vert^{2a(Q-a)}} 
\end{equation}
The explicit form of the function $d$ was found in 
ref. \cite{Fateev3}
\begin{subequations} \label{20}
\begin{eqnarray}
d(a\vert s) & = & \left[ {\pi z \Gamma(b^2) b^{2-2b^2} \over
\Gamma(1-b^2)} \right]^{(Q-2a)/2b} {G(Q-2a) \over G(2a-Q)}
\nonumber \\
& & \times \exp \left\{ - \int_{-\infty}^{\infty} {\rd t \over t}
\left[ {\sinh[(Q-2a)t] \cos^2(st/b) \over \sinh(bt) \sinh(b/t)}
- {(Q-2a) \over t} \right] \right\} \label{20a} \\
\ln G(x) & = & \int_0^{\infty} {\rd t \over t} \left[
{{\rm e}^{-Qt/2} - {\rm e}^{-xt} \over (1-{\rm e}^{-bt})
(1-{\rm e}^{-t/b})} + {(Q/2-x)^2 \over 2} {\rm e}^{-t}
+ {(Q/2-x) \over t} \right] \label{20b}
\end{eqnarray}
where $\Gamma$ denotes the Gamma function.
\end{subequations}
The auxiliary variable $s$ depends on $z$ and $z_B$ as follows
\begin{equation} \label{21}
\cosh^2(\pi s) = {z_B^2 \over z} \sin(\pi b^2)
\end{equation}
It is either real or pure imaginary.
Writing
\begin{equation} \label{22} 
s = u + {\rm i} v \quad \quad \quad u, v \in R
\end{equation}
one has namely
\begin{subequations}
\begin{eqnarray}
{z_B^2 \over z} \sin(\pi b^2) > 1 & & \quad \quad \quad
v=0, u\in (0,\infty) \label{23a} \\
{z_B^2 \over z} \sin(\pi b^2) < 1 & & \quad \quad \quad
u=0, v\in (0,1/2) \label{23b}
\end{eqnarray}
\end{subequations}
The function $d$ clearly fulfils the unitarity relation
\begin{equation} \label{24}
d(a \vert s) d(Q-a \vert s) = 1 
\end{equation}
It is straightforward to show from (\ref{19}) that the boundary
exponential operators exhibit the following reflection property
\begin{equation} \label{25}
{\rm e}^{a\phi_B}(y) = d(a\vert s) {\rm e}^{(Q-a)\phi_B}(y)
\end{equation}

The associated boundary sinh-Gordon model is defined by the action
\begin{equation} \label{26}
{\cal A}_{shG}(z,z_B)  =  \int_{x>0} {\rm d}^2 r \left[
{1\over 4 \pi} (\nabla \phi)^2 + 2 z \cosh (2 b \phi) \right]
+ 2 z_B \int_{-\infty}^{\infty} \rd y ~ \cosh (b \phi_B)
\end{equation}
The vacuum expectation value $\langle {\rm e}^{a\phi_B}\rangle_{shG}$
is conjectured \cite{Fateev1} -- \cite{Fateev3}
to satisfy a reflection relation similar to (\ref{25})
\begin{equation} \label{27}
\langle {\rm e}^{a\phi_B} \rangle_{shG} = 
d(a\vert s) \langle {\rm e}^{(Q-a)\phi_B} \rangle_{shG}
\end{equation}
This relation, together with the obvious symmetry
\begin{equation} \label{28}
\langle {\rm e}^{a\phi_B} \rangle_{shG} = 
\langle {\rm e}^{-a\phi_B} \rangle_{shG}
\end{equation}
determines the expectations up to a periodic function;
the solution of interest is a ``minimal solution'' to the
functional equations (\ref{27}) and (\ref{28}).

The boundary sine-Gordon model with the action (\ref{12b}) results
as an analytical continuation of (\ref{26}) via the
substitutions $b \to {\rm i}b, z\to -z, z_B\to -z_B$.
The final result for $\langle {\rm e}^{{\rm i}a\phi_B}\rangle_{sG}
\equiv \langle {\rm e}^{{\rm i}a\phi_B}\rangle_{z,z_B}$
reads [see eqs. (4.10) and (4.11) of ref. \cite{Fateev3}, add
a missing factor $1/2$ to $\ln g_S(a)$]
\begin{equation} \label{29}
\langle {\rm e}^{{\rm i}a\phi_B}\rangle_{z,z_B} =
\left[ {\pi z \Gamma(1-b^2) \over \Gamma(b^2)} 
\right]^{{a^2\over 2(1-b^2)}} g_0(a) g_S(a)
\end{equation}
where
\begin{subequations} 
\begin{eqnarray}
\ln g_0(a) & = & \int_0^{\infty} {\rd t \over t} \left\{
{2 \sinh^2(abt) \left[ {\rm e}^{(1-b^2)t/2} \cosh(t/2)
\cosh(b^2t/2)-1\right] \over \sinh t \sinh(b^2 t) 
\sinh((1-b^2)t)} - a^2 {\rm e}^{-t} \right\} \label{30a} \\
\ln g_S(a) & = & 
\int_0^{\infty} {\rd t \over t} 
{2 \sinh^2(abt) \sin^2(s t) \over \sinh t \sinh(b^2 t) 
\sinh((1-b^2)t)} \label{30b}
\end{eqnarray}
\end{subequations}
where we have taken into account that $s^* = u - {\rm i} v = \pm s$.
This result holds under the conformal normalization of the bulk
and boundary fields corresponding to the short distance asymptotics
\begin{subequations}
\begin{eqnarray}
{\rm e}^{2{\rm i}a\phi}({\vek r}){\rm e}^{-2{\rm i}a\phi}({\vek r}')
& = & {1\over \vert {\vek r} - {\vek r}' \vert^{4 a^2}} \quad \quad
{\rm as}\ \vert {\vek r} - {\vek r}' \vert \to 0 \label{31a} \\
{\rm e}^{{\rm i}a\phi_B}(y){\rm e}^{-{\rm i}a\phi_B}(y')
& = & {1\over \vert y - y' \vert^{2 a^2}} \quad \quad
{\rm as}\ \vert y - y' \vert \to 0 \label{31b}
\end{eqnarray}
\end{subequations}
\cite{Zamolodchikov}, \cite{Lukyanov}.
This normalization is consistent with a well known leading
short-distance behaviour of the positive-negative pair 
correlation in the Coulomb gas.
For the case of interest $a=b$, using the integral representation
\begin{equation} \label{32}
\ln \Gamma(x) = \int_0^{\infty} {\rd t \over t} {\rm e}^{-t}
\left[ x-1 + {{\rm e}^{-(x-1)t} - 1 \over 1 - {\rm e}^{-t}} \right],
\quad \quad {\rm Re} ~ x >0
\end{equation}
one gets after some algebra
\begin{subequations}
\begin{eqnarray}
g_0(b) & = & {1\over 2 \sqrt{\pi}(1-b^2)^2 \Gamma(b^2)} 
\Gamma \left( {1-2b^2\over 2-2b^2} \right)
\Gamma \left( {b^2\over 2-2b^2}\right) 
\label{33a} \\
g_S(b) & = & (1-b^2) {\sinh \left( {\pi s\over 1-b^2}\right) 
\over \sinh(\pi s)} \label{33b}
\end{eqnarray}
\end{subequations}
With the aid of (\ref{21}), one then finds
\begin{eqnarray} \label{34}
\langle {\rm e}^{{\rm i}b\phi_B} \rangle_{z,z_B} & = &
{1\over 4 \pi^{3/2} (1-b^2) z_B}
\Gamma \left( {1-2b^2\over 2-2b^2} \right)
\Gamma \left( {b^2\over 2-2b^2}\right) 
\left[ {2\pi z_B \over \Gamma(b^2)} \right]^{1/(1-b^2)}
\nonumber \\
& & \times \left[ {1\over 2\cosh (\pi s)} \right]^{b^2/(1-b^2)}
{\sinh \left( {\pi s\over 1-b^2}\right) 
\over \sinh(\pi s)}
\end{eqnarray}
According to (\ref{21}), in the limit $z\to 0$ the real parameter $s$
diverges and (\ref{34}) yields
\begin{equation} \label{35}
\langle {\rm e}^{{\rm i}b\phi_B} \rangle_{z=0,z_B}  = 
{1\over 4 \pi^{3/2} (1-b^2) z_B}
\Gamma \left( {1-2b^2\over 2-2b^2} \right)
\Gamma \left( {b^2\over 2-2b^2}\right) 
\left[ {2\pi z_B \over \Gamma(b^2)} \right]^{1/(1-b^2)}
\end{equation}
in full agreement with formula (24) in ref. \cite{Fateev1}.

\section{Surface tension}
We first summarize the bulk thermodynamics of the 2D Coulomb gas
\cite{Samaj1}.
The pressure is expressible in terms of the soliton mass
$M$ as follows
\begin{equation} \label{36}
\beta p = {M^2 \over 4} \tan \left( {q \pi \over 2} \right)
\end{equation}
where $q=\beta/(4-\beta)$.
The normalization of the bulk field (\ref{31a}) fixes the
relationship between the fugacity $z$ and $M$ in the form
\begin{equation} \label{37}
z = {\Gamma(q/(q+1)) \over \pi \Gamma(1/(q+1))} 
\left[ M {\sqrt{\pi} \Gamma((q+1)/2) \over 2 \Gamma(q/2)}
\right]^{2/(q+1)}
\end{equation}
The total particle number density $n$, generated via 
\begin{equation} \label{38}
n = z {\partial \over \partial z} \beta p
\end{equation}
is related to $M$ as follows
\begin{equation} \label{39}
n = {1\over 4} M^2 (1+q) \tan \left( {\pi q \over 2} \right)
\end{equation}

Let us denote the surface tensions of the metal-electrolyte
and of the dielectric-electrolyte boundaries by
\begin{subequations} \label{40}
\begin{eqnarray} 
\gamma_{met} & = & \lim_{z_B\to\infty} \gamma(z,z_B) \label{40a} \\
\gamma_{diel} & = & \lim_{z_B\to 0} \gamma(z,z_B) \label{40b}
\end{eqnarray}
\end{subequations}
respectively. 
In paper \cite{Samaj2} [formula (33b) with zero bulk potential
$\xi = 0$], we have found
\begin{equation} \label{41}
\beta \gamma_{met} = - {M \over 4 \cos(q\pi/2)} \left[ 
\sin \left( {q\pi \over 2} \right) - \cos \left( {q\pi \over 2} \right) 
+ 1 \right]
\end{equation}
With regard to the definitions (\ref{40}) and eq. (\ref{15}), it holds
\begin{eqnarray} \label{42}
\beta \gamma_{diel} & = & \beta \gamma_{met} - \int_0^{\infty}
\rd z_B {\partial \over \partial z_B} \beta \gamma(z,z_B)
\nonumber \\
& = & \beta \gamma_{met} + 2 \int_0^{\infty} \rd z_B \left[
\langle {\rm e}^{{\rm i}b\phi_B} \rangle_{z,z_B} -
\langle {\rm e}^{{\rm i}b\phi_B} \rangle_{z=0,z_B} \right]
\end{eqnarray}
Inserting the previously obtained formulae (\ref{34}) and (\ref{35})
into (\ref{42}) and substituting $b=\sqrt{\beta}/2$ in accordance
with (\ref{11c}), one arrives at
\begin{equation} \label{43} 
\beta \gamma_{diel} = \beta \gamma_{met} + {2\over \pi^{3/2} (4-\beta)}  
\left[ {2\pi \over \Gamma(\beta/4)} \right]^{4/(4-\beta)}
\Gamma \left( {2-\beta \over 4-\beta} \right) 
\Gamma \left( {\beta \over 8-2\beta} \right) ( I_1 + I_2 )
\end{equation}
where
\begin{subequations}
\begin{eqnarray}
I_1 & = & 2\pi \left[ {z\over 4 \sin(\pi\beta/4)} \right]^{2/(4-\beta)}
\nonumber \\ & & \times
\int_0^{\infty} \rd u \left\{ \sinh \left( {4\pi u \over 4-\beta} 
\right) - \sinh (\pi u) \left[ 2 \cosh (\pi u) \right]^{\beta/(4-\beta)}
\right\} \label{44a} \\
I_2 & = & 2\pi \left[ {z\over 4 \sin(\pi\beta/4)} \right]^{2/(4-\beta)}
\nonumber \\ & & \times
\int_0^{1/2} \rd v \left\{ \sin \left( {4\pi v \over 4-\beta} 
\right) - \sin (\pi v) \left[ 2 \cos (\pi v) \right]^{\beta/(4-\beta)}
\right\} \label{44b} 
\end{eqnarray}
\end{subequations}
Here, by using eq. (\ref{21}) we have transformed the integration
over $z_B$ into integrations over the  $u, v$-components of the
auxiliary variable $s$ (\ref{22}); the split of the integral
onto the $I_1$ and $I_2$ ones is due to the existence of the
two regimes (\ref{23a}) and (\ref{23b}).
Algebra gives
\begin{subequations} 
\begin{eqnarray}
I_1 & = & {1\over 2} (4-\beta) \left[ {z\over 4 \sin(\pi\beta/4)} 
\right]^{2/(4-\beta)} \left( 2^{\beta/(4-\beta)} - 1 \right) \label{45a} \\
I_2 & = & {1\over 2} (4-\beta) \left[ {z\over 4 \sin(\pi\beta/4)} 
\right]^{2/(4-\beta)} \left[ 1 - 2^{\beta/(4-\beta)} - \cos \left(
{2\pi \over 4-\beta} \right) \right] \label{45b}
\end{eqnarray}
\end{subequations}
Inserting $I_1$ and $I_2$ into (\ref{43}), then using (\ref{37}),
applying simple operations with trigonometric functions and formula
$\Gamma(x) \Gamma(1-x) = \pi / \sin(\pi x)$ and considering (\ref{41}),
one finally gets
\begin{equation} \label{46}
\beta \gamma_{diel} =  {M \over 4 \cos(q\pi/2)} \left[ 
\sin \left( {q\pi \over 2} \right) + \cos \left( {q\pi \over 2} \right) 
- 1 \right]
\end{equation}
where $q=\beta/(4-\beta)$ and $M$ is related to the fugacity $z$
by (\ref{37}).

When $M$ is expressed in terms of the particle density $n$ by using 
eq. (\ref{39}), (\ref{46}) takes the form
\begin{equation} \label{47}
\beta \gamma_{diel} = {1\over 2} \left[ {n (4-\beta) \over 
2 \sin(\pi\beta/(4-\beta))} \right]^{1/2} 
\left[ \sin \left( {\pi\beta \over 2(4-\beta)} \right) + 
\cos \left( {\pi \beta \over 2(4-\beta)} \right) - 1 \right]
\end{equation}
$\beta\gamma_{diel}$ has the small $\beta$-expansion
\begin{equation} \label{48}
\beta \gamma_{diel} = {1\over 8} (2\pi\beta n)^{1/2}
\left[ 1 - {\pi \over 16} \beta - {\pi \over 64} \left(
1-{\pi \over 6} \right) \beta^2 + \ldots \right]
\end{equation}
The singular behaviour of $n$ as $\beta \to 2^-$ ($z$ fixed)
can be deduced from formulae (\ref{37}) and (\ref{39}):
\begin{equation} \label{49}
n \sim {4 z^2 \pi \over 2-\beta}
\end{equation}
On the other hand, the surface tension, expressed in terms
of the fugacity $z$, remains finite 
at the bulk collapse point $\beta = 2$,
\begin{equation} \label{50}
\gamma_{diel} = {\pi z \over 4}, \quad \quad \quad \beta = 2
\end{equation}
The function $\beta \gamma_{diel}$ increases monotonously
up to $\beta = 3$ where the surface tension diverges,
\begin{equation} \label{51}
\gamma_{diel} \sim {\pi \over 12} \left[ {\Gamma(1/4)
\over \Gamma(3/4)} \right]^2 {z^2 \over 3-\beta},
\quad \quad \quad  \beta \to 3^-
\end{equation} 
The singularity prevents from going beyond this point.

\section{Analytic checks}
The surface tension $\gamma$ $(=\gamma_{diel})$ is the boundary
part per unit length of the grand potential $\Omega$.
The total number of particles is given by
$N = N_+ + N_- = - \beta z \partial \Omega / \partial z$.
The boundary part of this relation is
\begin{equation} \label{52}
-\beta z {\partial \gamma \over \partial z} =
\int_0^{\infty} \rd x ~ [n(x) - n]
\end{equation}
or, since $z \propto n^{1-\beta/4}$,
\begin{equation} \label{53}
-\beta n {\partial \gamma \over \partial n} = \left( 1-
{\beta \over 4} \right) \int_0^{\infty} \rd x ~ [n(x) - n]
\end{equation}
These formulae will be used for computing the surface tension
from the density profile.

\subsection{High-temperature expansion}
As a check of the exact expression (\ref{47}) for the surface
tension, its high-temperature expansion in $\beta$, equation (\ref{48}),
can be compared to a direct evaluation of the first two terms
of this expansion by using a renormalized Mayer expansion,
in close analogy with section 5 of paper \cite{Samaj2}).
Since the derivative of the interaction potential 
$v({\vek r},{\vek r}')+v({\vek r}^*,{\vek r}')$
with respect to $x$ vanishes when point ${\vek r}$ is 
on the interface, the same boundary condition holds for 
the renormalized bond $K({\vek r},{\vek r}')$,
\begin{equation} \label{54}
{\partial \over \partial x} K({\vek r},{\vek r}') \big\vert_{x=0} = 0
\end{equation}
After some algebra, one finds
\begin{equation} \label{55}
\int_0^{\infty} \rd x \left[ n(x) - n \right] = 
{(2\pi\beta n)^{1/2} \over 16}
\left[ - 1 - {\beta \over 4} \left( 1 + {\pi \over 4} \right)
+ {\beta \pi \over 8} + \ldots \right]
\end{equation}
Consideration of (\ref{55}) in (\ref{53}) gives the final result
\begin{equation} \label{56}
\beta \gamma = {1\over 8} (2\pi \beta n)^{1/2} \left[
1 - {\pi \over 16} \beta + O(\beta^2) \right]
\end{equation}
in agreement with (\ref{48}).

\subsection{Free fermion $\beta=2$ case}
Using a Thirring-like representation of the Coulomb gas in presence
of a dielectric wall at special inverse temperature $\beta = 2$,
the exact result for the density profile was obtained in the form
\cite{Jancovici1}
\begin{equation} \label{57}
n_{\pm}(x) - n_{\pm} = - {m^2\over 2\pi} K_0(2mx)
\end{equation}
where $m = 2\pi z$ is the rescaled fugacity.
Thus,
\begin{equation} \label{58}
\int_0^{\infty} \rd x [ n(x) - n] = - {\pi z \over 2}
\end{equation}
Inserting this integral into equation (\ref{52}), one
immediately arrives at the exact formula (\ref{50}).

\section{Concluding remarks}
A two-dimensional model for the interface, the two-component 
plasma bounded by a rectilinear ideal dielectric wall, 
was considered.
The surface tension as a function of fugacity is finite
at any inverse temperature $\beta < 3$.
In general, the surface properties are supposed to be 
governed by the particle-image interaction.
A particle at distance $x$ from the dielectric wall
interacts with its own image of the same sign through a
repulsive potential $-(1/2)\ln(2x)$, and the corresponding
Boltzmann factor $(2x)^{\beta/2}$ is always integrable
at small $x$, and gives rise to a vanishing particle density
at the interface.
In spite of this, our exact solution for the surface tension
predicts a surface collapse at $\beta = 3$, identified with 
the divergence of the surface tension with a characteristic
type of short-distance collapse singularity, see formula
(\ref{51}). 
Such a phenomenon might be explained by recalling an observation
of Hansen and Viot \cite{Hansen} concerning the short-distance
behavior of bulk pair distribution functions for two charges 
of the same, let us say plus, sign.
The expected dependence $g_{++} \sim C_{++} r^{\beta}$
as $r\to 0$, where the prefactor $C_{++}$ is related to a
difference of free energies, is changed to $g_{++} \sim {\bar C}_{++} 
r^{2-\beta}$ at $\beta \ge 1$ as a consequence of the divergence of 
$C_{++}$ at point $\beta = 1$.
The weaking of $g_{++}$ at short distance is caused by
a pair formation of oppositely charged particles at low
temperatures: the neutrality of a pair allows a third
particle to approach very close to the pair.
For $\beta \ge 2$, the strong clustering of positive-negative
particles causes an effective short-distance attraction 
between particles of the same sign.
Based on these plausible arguments it is tempting to 
conjecture that, at $\beta = 3$, the surface tension diverges
due to a paradoxical short-distance collapse of a
particle with its own image of the same sign.

\section*{Acknowledgments}
I am grateful to Bernard Jancovici for stimulating discussions,
useful comments, and careful reading of the manuscript.
My stay in LPT Orsay is supported by a NATO fellowship.
A partial support by Grant VEGA 2/7174/20 is acknowledged.

\end{document}